\documentclass{emulateapj}
\usepackage{apjfonts}

\usepackage{subfigure}    

\newcommand{\Msun}      {\mbox{$\rm\,M_{\mathord\odot}$}}

\begin{document}

\submitted{To appear in the Astrophysical Journal}

\def\lsim{\mathrel{\lower .85ex\hbox{\rlap{$\sim$}\raise
.95ex\hbox{$<$} }}}
\def\gsim{\mathrel{\lower .80ex\hbox{\rlap{$\sim$}\raise
.90ex\hbox{$>$} }}}

\title{{\em XMM-Newton} Finds That SAX J1750.8$-$2900 May Harbor the Hottest, Most Luminous Known Neutron Star}

\author{A. W. Lowell\altaffilmark{1},
J. A. Tomsick\altaffilmark{1},
C. O. Heinke\altaffilmark{2},
A. Bodaghee\altaffilmark{1},
S. E. Boggs\altaffilmark{1},
P. Kaaret\altaffilmark{3},
S. Chaty\altaffilmark{4},
J. Rodriguez\altaffilmark{4},
R. Walter\altaffilmark{5}}

\altaffiltext{1}{Space Sciences Laboratory, 7 Gauss Way, 
University of California, Berkeley, CA 94720-7450, USA
(e-mail: alowell@ssl.berkeley.edu)}

\altaffiltext{2}{Department of Physics, University of Alberta,
Room 238 CEB, Edmonton, AB T6G 2G7, Canada}

\altaffiltext{3}{Department of Physics and Astronomy, University of
Iowa, Iowa City, IA 52242, USA}

\altaffiltext{4}{AIM - Astrophysique Instrumentation Mod\'elisation
(UMR 7158 CEA/CNRS/Universit\'e Paris 7 Denis Diderot),
CEA Saclay, DSM/IRFU/Service d'Astrophysique, B\^at. 709,
L'Orme des Merisiers, FR-91 191 Gif-sur-Yvette Cedex, France}

\altaffiltext{5}{INTEGRAL Science Data Centre, Observatoire
de Gen\`eve, Universit\'e de Gen\`eve, Chemin d'Ecogia, 16,
1290 Versoix, Switzerland} 

\begin{abstract}

We have performed the first sensitive X-ray observation of the low-mass X-ray binary (LMXB) SAX J1750.8$-$2900 in quiescence with {\em XMM-Newton}.
The spectrum was fit to  both a classical black body model, and a non-magnetized, pure hydrogen neutron star (NS) atmosphere model.  A power law component was
added to these models, but we found that it was not required by the fits.  The distance to SAX J1750.8$-$2900 is known to be D = 6.79 kpc from a previous analysis of
photospheric radius expansion bursts.  This distance implies a bolometric luminosity (as given by the NS atmosphere model) of $(1.05 \pm 0.12)$ $\times$ 10$^{34}$ (D/6.79 kpc)$^{2}$ erg s$^{-1}$, which is the 
highest known luminosity for a NS LMXB in quiescence.  One simple explanation for this surprising result could be that the crust and core of the NS were not in thermal 
equilibrium during the observation.  We argue that this was likely not the case, and that the core temperature of the NS in SAX J1750.8$-$2900 is unusually high.

\end{abstract}


\section{Introduction}

Transiently accreting neutron stars in low-mass X-ray binary systems accrete
matter from a $\la 1 \Msun$ donor star for weeks to months before returning to 
quiescence.  During an accretion episode, matter falls onto the surface of the NS and is strongly compressed
due to the intense gravity.  Pycnonuclear reactions occur in the resulting high-density matter \citep{Haensel1990} and proceed to heat the crust out of
thermal equilibrium with the core.  When the 
system returns to quiescence, the crust begins to thermally relax by conducting
a fraction of the excess heat into the core until equilibrium is reestablished.
This fraction depends on the thermal conductivity of the crust \citep{Shternin2007,Brown2009}, the amount of hydrogen and helium remaining
in the NS atmosphere post-outburst \citep{Brown2002}, and the temperature of the core.
The remaining energy in the crust is believed to be thermally radiated away.

The X-ray spectrum of this thermal radiation differs significantly from a typical 
black body spectrum due to the atmosphere of the NS, and to the 
intense gravity at the NS surface \citep{Zavlin1996}.  The surface temperature may be deduced from 
spectral fits to a physical model which accounts for the previously mentioned 
effects.  In thermal equilibrium, the temperature of the surface will be equal to 
that of the core.  Thus, X-ray observations of quiescent NS LMXB systems for
which the crust and core have reached thermal equilibrium yield measurements
of the NS core temperature.  The temperature of the core does not change appreciably
over timescales $<10^{4}$ years and is believed to be set by both the mass accretion history
of the system \citep{Brown1998} and the efficiencies of the neutrino producing
mechanisms in the core.  The core temperatures and bolometric luminosities for these 
sources are very interesting parameters as they constrain neutrino emission models 
and give insight into the neutron degenerate matter equation of state.
 
Here we present the first X-ray observation of the LMXB NS transient SAX J1750.8$-$2900 in quiescence.
SAX J1750.8$-$2900 (hereafter referred to as J1750) was first detected by the Wide Field Cameras 
(WFCs) aboard the {\em BeppoSAX} satellite in 1997 \citep{Natalucci1999}.  The WFCs detected nine 
separate Type I X-ray bursts from J1750  with intensities ranging from 0.4 to 1.0 Crab. From these bursts, 
a 3 $\sigma$ upper limit of $\sim$ 7 kpc was estimated for the source distance. A separate
observation by \citet{Kaaret2002} in 2001 with the {\em Rossi X-ray Timing Explorer's} ({\em RXTE}) Proportional
Counter Array (PCA) revealed Type I X-ray bursts accompanied by millisecond quasiperiodic oscillations, allowing
the authors to estimate the NS spin to be 601 Hz.  Two of the four bursts from the 2001 {\em RXTE}/PCA observation were
found to display evidence of photospheric radius expansion (PRE) \citep{Kaaret2002,Galloway2008} thus allowing for an
estimation of the distance to the source.  \citet{Galloway2008} suggest $D = 6.79 \pm 0.14$ kpc for H-poor burning, and
$D = 5.21 \pm 0.11$ kpc for H-rich burning.  The four observed bursts were found to have decay times of $\tau = 5 - 7.3$ s, which
\citet{Galloway2008} suggest is indicative of H-poor burning.  Thus, we assume for the remainder of this work that the distance
to J1750 is $6.79 \pm 0.14$ kpc.

Since its discovery, J1750 has been reported to be in outburst four times.  The outbursts in 2001 and 2008 \citep{Kaaret2002,Markwardt2008} were of relatively long duration
(4 - 5 months), while the outbursts in 1997 and 2011 \citep{Natalucci1999,Kuulkers2011,Natalucci2011} were relatively short 
($\le$ 1 month).  The {\em RXTE} All-Sky Monitor (ASM) light curve shown in the top panel of Figure 1 clearly depicts the outburst from 2008 along with a square denoting the time of our {\em XMM}
observation of J1750 in 2010.  The bottom panel shows the {\em RXTE}/PCA light curve over the same time period.  As we show in Section 4, the spectrum of J1750 was found to be described
purely by a thermal component, which, along with the lack of X-ray activity in the {\em RXTE}/ASM and {\em RXTE}/PCA lightcurves around April 7, 2010, strongly suggests that J1750 was in quiescence during our observation with {\em XMM}.

\begin{figure}
\centering
\includegraphics[width=0.48\textwidth]{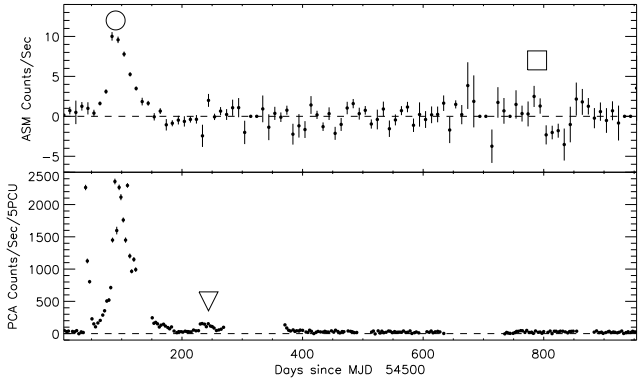}
\caption{The top panel shows the 1.5$-$12 keV, Daily-averaged {\em RXTE}/ASM light curve of J1750 with 10-day bins, while the bottom panel is the 2$-$10 keV {\em RXTE}/PCA light curve.  The last major outburst to have occurred before our 
observation was in 2008 and is denoted by the circle.  The time of our {\em XMM-Newton} observation is marked by the square at MJD 55293.  The detection of a flare by {\em Swift} \citep{Linares2008b} is denoted by the triangle.  The {\em RXTE}/PCA data gap
between MJD $\sim$ 54769$-$54870 is bracketed by a rise and a fall in the X-ray count rate, which suggests that this time period may have contained undetected X-ray activity.   For the {\em RXTE}/PCA light curve, the plotting symbols are 
larger than the error bars for most of the data.}
\end{figure}

\section{Observations and Analysis}

The field containing J1750 was observed on April 7, 2010, from 11:21:08 
to 17:37:45 UT using the pn, MOS1, and MOS2 instruments aboard the {\em XMM-Newton} telescope 
(Observation ID 0603850201, Revolution Number 1891) with a medium filter.  J1750 was located approximately 7$^{\prime}$ 
off-axis since it was not the primary target of the observation.  The total
exposure time was 21.8 ks.  However, a light-curve of the observation revealed
that there was strong background flaring for the final 1.9 ks of the observation, so only 
the first 19.9 ks were considered in the analysis that follows. The total livetime 
was 16.0 ks for the pn detector, and 19.6 ks for the MOS1 and MOS2 detectors.

We performed the data analysis with the {\em XMM-Newton} Science Analysis System (SAS).  The tasks
{\ttfamily emchain} and {\ttfamily epchain} were used to produce event lists for the pn, MOS1, and MOS2 
detectors.  A temporal filter was applied to the event lists using the task {\ttfamily evselect} 
in order to remove the period that contained the aforementioned flaring.

The source spectra were extracted (using the task {\ttfamily evselect}) from appropriately sized (25$^{\prime\prime}$ 
radius for pn, 30$^{\prime\prime}$ for MOS1, and 30$^{\prime\prime}$ for MOS2)  circular regions centered 
about the observed position of J1750, which was found to be consistent with the {\em Chandra} position given in \citet{Chakrabarty2008}.
The total number of counts (after background subtraction) was 269 for the pn detector, 153 for the MOS1 detector, and 112 for the MOS2 detector.
We applied a KS test to the event lists from the source extraction regions and found no evidence of variability. 
Our procedure for choosing the background spectral extraction regions was slightly
different for each detector due to the proximity of J1750 to straylight artifacts in the data. 
The straylight contamination appears as circular arcs centered about a point outside of the field of view, and is likely due
to a bright source located outside of the field of view. In the pn and especially the MOS1 detectors, the
position of J1750 coincided with a straylight arc, and thus the background regions for these
detectors were chosen so as to include part of the same streak. For the MOS2 detector,
we chose a large, circular, source-free region from which to sample the background, since
the position of J1750 was not found to coincide with a streak in the MOS2 data.  Figure 2 shows the {\em XMM} images
of J1750 for all three detectors, along with the extraction regions and the straylight contamination.

We used {\ttfamily rmfgen} to create response matrices and {\ttfamily arfgen} to create ancillary response files. The fits were performed on the raw, unbinned spectra using
W statistics \citep{Cash1979}, due to the small number of counts in the spectrum. The spectra were loaded into XSPEC V12 \citep{Arnaud1996} and fit simultaneously over the range 0.3$-$12.0 keV with
all physical parameters tied together.  We used the abundances from \citet{Wilms2000} and photoelectric cross-sections from \citet{Balucinska1992} and \citet{Yan1998}.  
The models used for the spectral fits were multiplied by an instrumental cross-calibration coefficient
which was held fixed at $C_{\rm pn}$ = 1.0 for the pn spectrum, but left free for the MOS1 and MOS2 spectra ($C_{\rm MOS}$).  This
parameter gave an indication of how well the MOS1 and MOS2 detectors agreed with the pn detector.  The best-fit values for this parameter can 
be found in Table 1.

\begin{figure}[ht!] 
\label{fig:subfigures}
\begin{center}
\label{fig:first}
\subfigure[]{\includegraphics[width=0.3\textwidth]{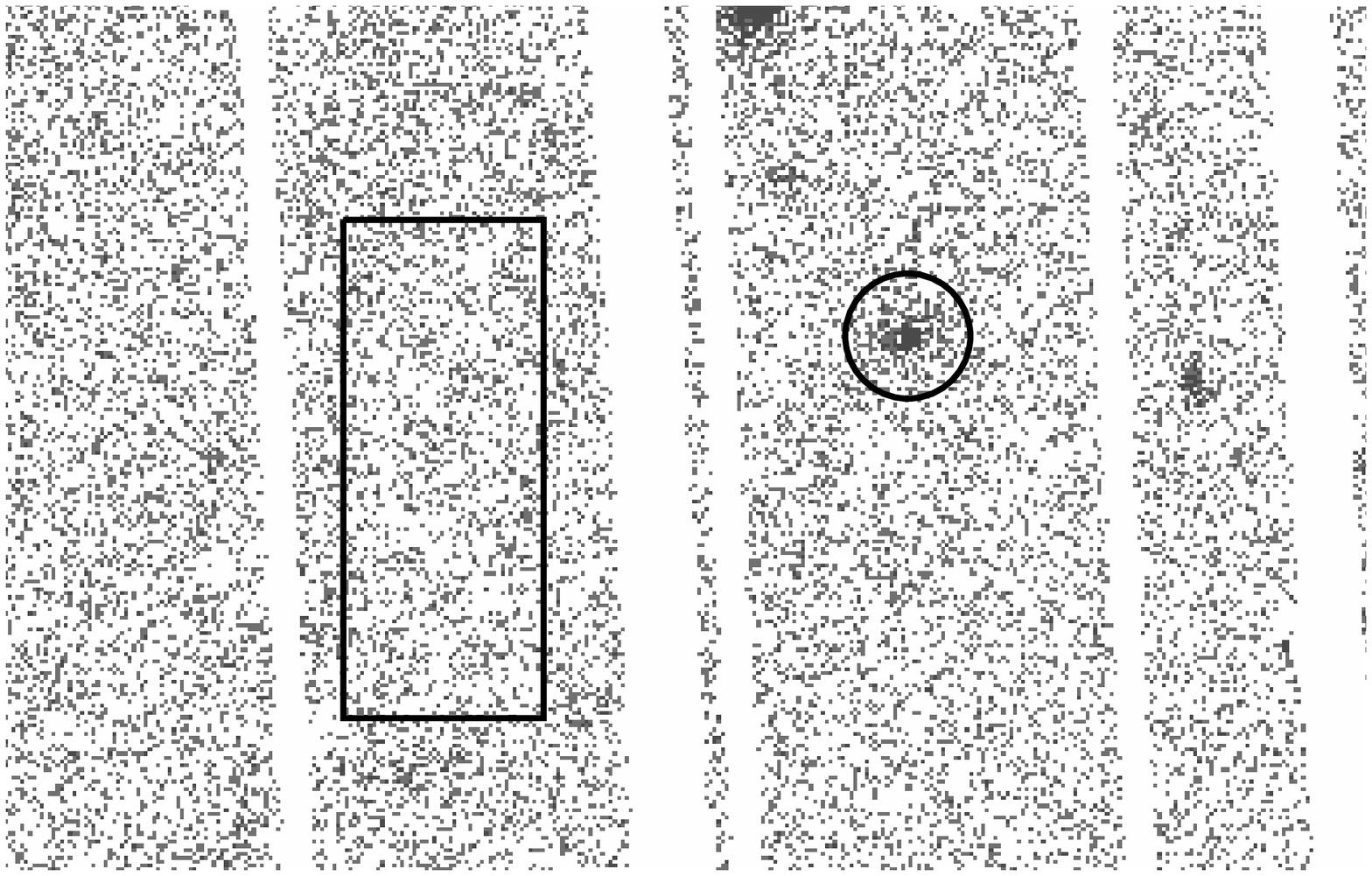}}
\label{fig:second}
\subfigure[]{\includegraphics[width=0.3\textwidth]{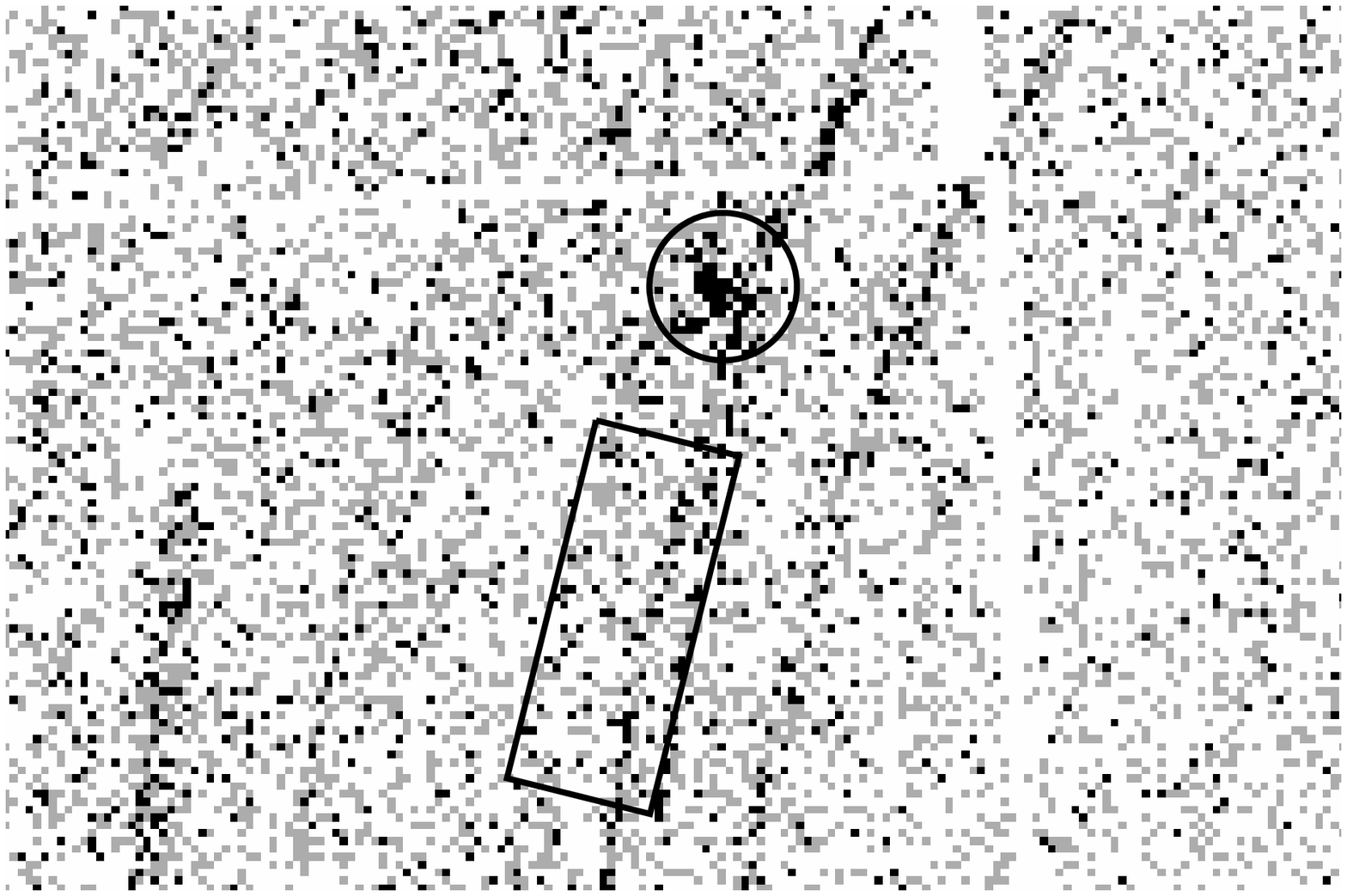}}
\label{fig:third}
\subfigure[]{\includegraphics[width=0.3\textwidth]{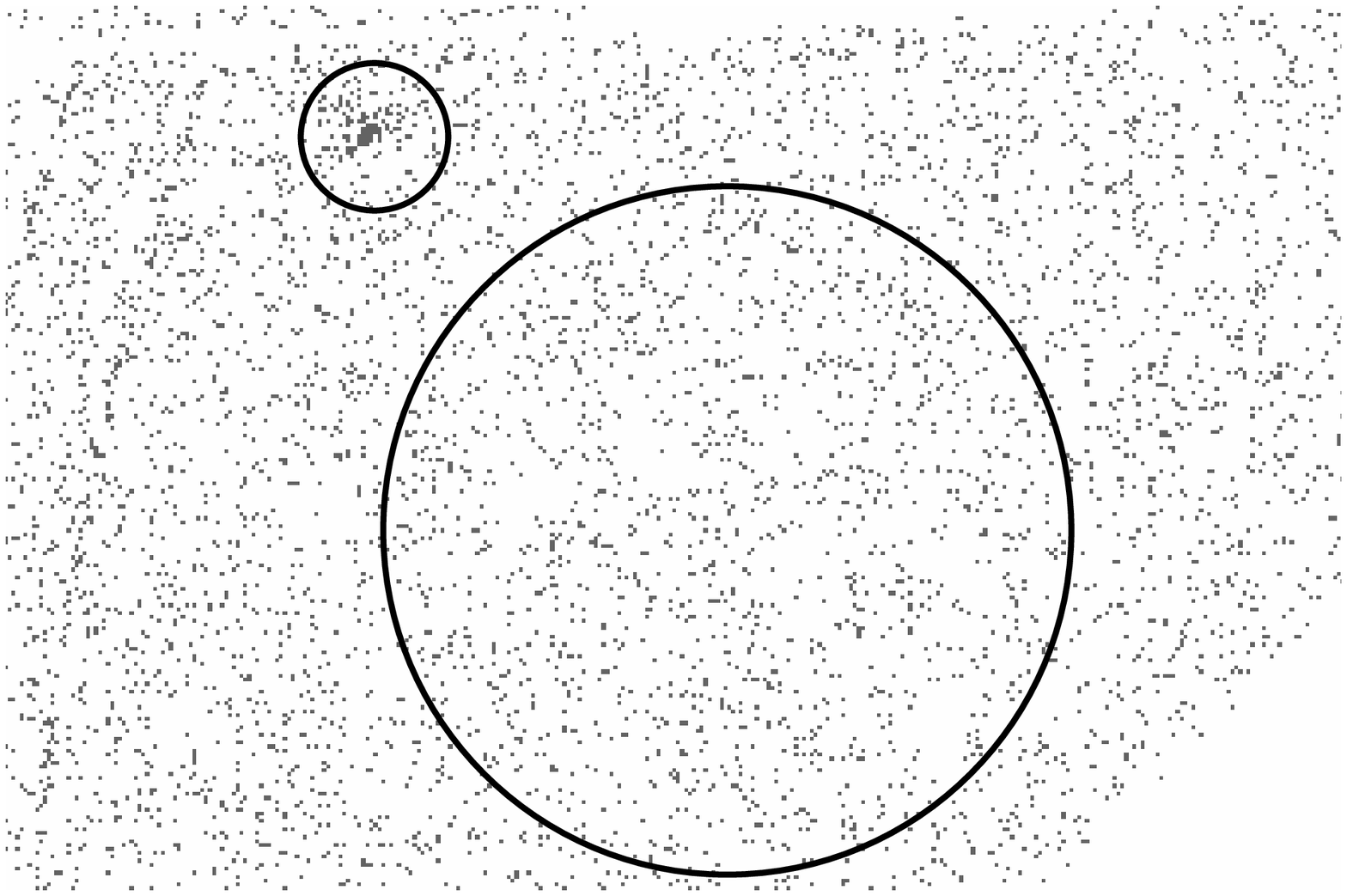}}
\end{center}
\caption{0.3 $-$ 12 keV images of J1750 with source and background extraction regions for the (a) pn, (b) MOS1, and (c) MOS2 detectors.  Straylight contamination is clearly visible in the MOS1 image.}
\end{figure}

\section{Results}

An initial fit to an absorbed power law model yielded a photon index of $\Gamma = 6.0^{+1.0}_{-0.9}$, a column density of 
$N_{\rm H} =$ 7.9$^{+1.7}_{-1.5}$ $\times$ 10$^{22}$ cm$^{-2}$, and a 0.5$-$10 keV absorbed flux of (1.2 $\pm$ 0.1) $\times$ $10^{-13}$ erg cm$^{-2}$ s$^{-1}$.  Note that this high column density is a 
result of the steep photon index, as the two are necessarily correlated for a simple, absorbed power law model.  Nonetheless, the steep photon index given by this fit suggests that the spectrum is best described by a thermal model.
Table 1 summarizes the results of our spectral fits for both a simple blackbody model ({\ttfamily bbodyrad}), and for a non-magnetized, pure hydrogen NS atmosphere
model ({\ttfamily nsatmos}).  All physical parameters were left free in the {\ttfamily bbodyrad} fits. In the {\ttfamily nsatmos} fits, the free parameters were the
absorption due to hydrogen gas $N_{\rm H}$ and the effective temperature\footnote{The effective temperatures for the {\ttfamily bbodyrad} and {\ttfamily nsatmos} models 
were multiplied by the gravitational redshift parameter $g_{\rm r}$ in order to determine the temperature as measured by an observer at infinity.  For a NS mass 
of 1.4 $\Msun$ and a NS radius of 10 km, $g_{\rm r}$ = 0.77.} of the surface $kT^{\infty}_{\rm eff}$.  The NS mass and radius were fixed at the commonly used values of 1.4 
$\Msun$ and 10 km, respectively.  The {\ttfamily nsatmos} model contains a normalization parameter $K$ that controls how much of the NS surface is emitting.  
We kept $K$ fixed at 1 under the assumption that the emission was isotropic. The source distance parameter was held fixed at 6.79 kpc (see Section 1).  Using this distance,
and the fluxes from Table 1, we compute an unabsorbed,  0.5$-$10 keV luminosity of $2.8^{+0.2}_{-1.4}$ $\times$ 10$^{33}$ erg s$^{-1}$ for the {\ttfamily bbodyrad} 
model and $(8.8 \pm 1.2)$ $\times$ 10$^{33}$ erg s$^{-1}$ for the {\ttfamily nsatmos} model. A binned spectrum of J1750 and the best fit to the {\ttfamily nsatmos} 
model is shown in Figure 3.

\begin{figure}
\centering
\includegraphics[angle=-90,width=0.48\textwidth]{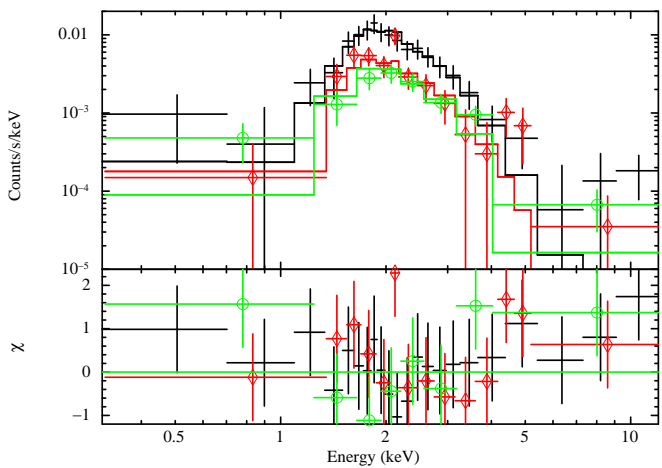}
\caption{0.3 $-$ 12 keV spectrum of J1750 for the pn (black), MOS1 (red, diamonds), and MOS2 (green, circles) detectors.  The best fit to the {\ttfamily nsatmos} model is shown.  Each
bin has a minimum of 20 counts.  The models were fit to the unbinned spectrum; this binned spectrum is for display purposes only.}
\end{figure}  


Spectra from NS LMXB transients are often found to display a power law component with $\Gamma \sim$ 1 - 2 \citep{Campana1998}. 
The reason for this is still not known, but it has been hypothesized that accretion at a low rate \citep{Zampieri1995} or a pulsar wind mechanism 
\citep{Illarionov1975,Stella1986,Campana2000} would produce hard X-rays resulting in a power law component. 
We searched for such a hard component in the spectrum by supplementing both the {\ttfamily bbodyrad} and {\ttfamily nsatmos} models with a power law and 
holding the photon index fixed at $\Gamma$ = 1 and $\Gamma$ = 2.  The results of these fits can be seen in the second and third columns of Table 1. In all cases, the best fit was 
found to have a value of zero for the power law normalization parameter, indicating that a power law was not required.  We calculated the maximum 
contribution of the power law flux to the total flux by fixing the power law normalization to its upper limit (90\% c.l.) and then refitting with all of the other parameters left free. The ratio of the power law 
normalization upper limit to the flux from the resulting fit is quoted as ``Power Law Contribution" in Table 1. 

\begin{table*}
\caption{Results from Spectral Fitting \label{tab:fits}}
\begin{minipage}{\linewidth}
\begin{center}
\footnotesize
\begin{tabular}{lccc} \hline \hline
Parameter & No Power Law & $\Gamma$ = 1 & $\Gamma$ = 2 \\ \hline

\multicolumn{4}{c}{nsatmos\footnote{For these fits, the NS mass, radius, and distance were fixed at 1.4 $\Msun$, 10 km, and 6.79 kpc respectively. Also, the normalization parameter K was held fixed under the assumption that the emission from the NS was isotropic.}}   \\ \hline
$N_{\rm H} (10^{22}$ cm$^{-2})$ & $5.9 \pm 0.5$ & $5.9 \pm 0.5$ & $5.8^{+0.6}_{-0.5}$ \\
$kT_{\rm eff}^{\infty}$ (eV) & $148 \pm 4$ & $149^{+4}_{-6}$ & $148 \pm 4$ \\
Power Law Norm. \footnote{Unabsorbed, 0.5 $-$ 10 keV power law flux in units of 10$^{-14}$ erg cm$^{-2}$ s$^{-1}$.} & - & $0.0^{+1.0}_{-0.0}$ & $0.0^{+2.5}_{-0.0}$ \\  
$C_{\rm MOS}$ & $1.2 \pm 0.2$ & $1.2 \pm 0.2$ & $1.2 \pm 0.2$ \\
C statistic value (dof) & 2047 (3897) & 2047 (3896) & 2047 (3896) \\
$F_{\rm bolo}$ (erg cm$^{-2}$ s$^{-1}) $\footnote{Unabsorbed, bolometric flux (0.01 $-$ 10 keV) on the pn detector for the best fit model.  Note that this may be interpreted as the thermal, bolometric flux, since the best fit value for the power law normalization parameter was zero in all cases.} & $ (1.9 \pm 0.2)$ $\times$ 10$^{-12}$ & $1.9^{+1.4}_{-0.1}$ $\times$ 10$^{-12}$ & $1.9^{+1.4}_{-0.1}$ $\times$ 10$^{-12}$ \\
$F_{0.5-10\rm keV}$ (erg cm$^{-2}$ s$^{-1})$\footnote{Unabsorbed, 0.5 $-$ 10 keV flux on the pn detector after refitting with the power law normalization parameter fixed at its upper limit.} & $(1.6 \pm 0.2)$ $\times$ 10$^{-12}$ & $(1.6 \pm 0.2)$ $\times$ 10$^{-12}$ & $(1.6 \pm 0.2)$ $\times$ 10$^{-12}$ \\ 
Power Law Contribution\footnote{Maximum possible contribution to the total 0.5 $-$ 10 keV flux from the power law component.} & - & 0.6\% & 1.6\% \\ \hline 

\multicolumn{4}{c}{bbodyrad} \\ \hline
$N_{\rm H} (10^{22}$ cm$^{-2})$ & $4.0^{+1.1}_{-0.9}$ & $4.0^{+1.0}_{-0.9}$ & $4.0^{+1.0}_{-0.9}$ \\
$kT_{\rm eff}^{\infty}$ (eV) & $331^{+43}_{-40}$ & $331^{+44}_{-40}$ & $331^{+44}_{-40}$ \\
Black Body Norm.\footnote{Black body normalization in units of (R/D)$^{2}$ where R is the source radius in km and D is the distance to the source in units of 10 kpc.} & $1.4^{+2.4}_{-0.8}$ & $1.4^{+2.3}_{-0.8}$ & $1.4^{+2.2}_{-0.8}$ \\
Power Law Norm. & - & $0.0^{+0.8}_{-0.0}$ & $0.0^{+2.0}_{-0.0}$ \\
$C_{\rm MOS}$ & $1.2 \pm 0.2$ & $1.2 \pm 0.2$ & $1.2 \pm 0.2$ \\
C statistic value (dof) & 2044 (3896) & 2044 (3895) & 2044 (3895) \\
$F_{\rm bolo}$ (erg cm$^{-2}$ s$^{-1})$ & $5.2^{+0.4}_{-2.4}$ $\times$ 10$^{-13}$ & $5.3^{+5.1}_{-5.3}$ $\times$ 10$^{-13}$ & $5.3^{+6.4}_{-4.0}$ $\times$ 10$^{-13}$ \\
$F_{0.5-10\rm keV}$ (erg cm$^{-2}$ s$^{-1})$ & $5.0^{+0.4}_{-2.5}$ $\times$ 10$^{-13}$ & $5.2^{+0.4}_{-2.7}$ $\times$ 10$^{-13}$ & $5.3^{+0.4}_{-2.3}$ $\times$ 10$^{-13}$ \\ 
Power Law Contribution & - & 1.7\% & 3.7\% \\ \hline 
\multicolumn{4}{l}{All errors are quoted at the 90\% confidence interval.} \\ \\
\end{tabular}
\end{center}
\end{minipage}
\end{table*}

\section{Discussion}

We have observed the transient NS LMXB J1750 in quiescence and fit its spectrum to both a black body and pure hydrogen NS
atmosphere model, with an additional power law component.  All of the fits yielded comparable values for the W statistic. 
The {\ttfamily nsatmos} model gives a 0.5$-$10. keV luminosity of $(8.8 \pm 1.2)$ $\times$ 10$^{33}$ erg s$^{-1}$, and
a bolometric luminosity of $(1.05 \pm 0.12)$ $\times$ 10$^{34}$ erg s$^{-1}$.  Our measured value of $kT^{\infty}_{\rm eff}$ = 148 $\pm$ 4 eV
for the NS surface temperature is higher than that seen for the four, quasi-persistent NS LMXB systems where a quiescent base temperature has been identified: $\sim$ 123 eV 
for XTE J1701$-$462 \citep{Fridriksson2011}, $<$ 70 eV for KS 1731$-$260 \citep{Cackett2010_ks}, 54 $\pm$ 2 eV for MXB 1659$-$29 \citep{Cackett2008}, and 109.4 $\pm$ 2.0 eV for
EXO 0748$-$676 \citep{Degenaar2011_2}.

A comparison between J1750 and other quiescent LMXB (qLMXB) systems
\citep[see][]{Tomsick2005,Heinke2010} suggests that J1750 could be the most luminous known NS LMXB in quiescence.  Figure 5 of \citet{Tomsick2005} displays a list of transient X-ray systems, both NSs and 
black holes (BHs), ordered by their orbital periods with their Eddington-scaled luminosities log$_{10}$($L_{\rm min}$/$L_{\rm Edd}$) plotted along the horizontal axis.  The
0.5$-$10 keV luminosity gives log$_{10}$($L_{\rm min}$/$L_{\rm Edd}$) = $-$4.3 for J1750, which is higher than any other qLMXB system listed\footnote{At first glance, J1750 appears to be less 
luminous than EXO 0748$-$676.  However, the analysis of the quiescent observation \citep{Garcia1999} of EXO 0748-676 with the {\em Einstein X-ray Observatory}, which gave a luminosity of 
$1.0^{+0.5}_{-0.2}$ $\times$ $10^{34}$ erg s$^{-1}$, assumed a source distance of 10 kpc, whereas the distance to the source is now estimated to be $\sim$ 7.4 kpc \citep{Galloway2008}.  
Additionally, a current, detailed study of EXO 0748$-$676 by \citet{Degenaar2011_2}
indicates that the quiescent, bolometric luminosity is actually $(6.0 \pm 0.2)$ $\times$ 10$^{33}$ erg s$^{-1}$.}.  The median, Eddington-scaled luminosities are shown seperately 
for NS and BH populations, and it is evident that the BH systems are generally less luminous.  One explanation for this discrepancy is that a portion of the accretion power 
is lost when matter crosses the event horizon of a black hole \citep{Narayan1997}.  Another explanation put forth by \citet{Fender2003} is that at low mass accretion rates, the power output of
black holes may be dominated by outflows of particles in jets rather than radiation.  The addition of J1750 to the NS population would raise the median
Eddington-scaled luminosity and thus increase this discrepancy. 

Figure 8 of \citet{Heinke2010} displays the quiescent, bolometric luminosities for various NS systems as a function of their time-averaged mass transfer rate.  The cooling curves
of \citet{Yakovlev2004} for various cooling scenarios are plotted alongside these sources.  In order to estimate the time-averaged mass transfer rate for J1750, we first
calculated the time-averaged luminosity by finding the average 1.5$-$12 keV count rate during outburst (as given by {\em RXTE}/ASM), and converting this to a 0.1$-$20 keV flux
using PIMMS\footnote{The Portable, Interactive, Multi-Mission Simulator (PIMMS) is used to estimate fluxes given the count rate and spectral shape as measured by a particular instrument, or to convert count rates between instruments.  We used the
web interface for PIMMS located at http://heasarc.nasa.gov/Tools/w3pimms.html.}, assuming a power law spectral shape with $\Gamma$ = 2 and $N_{H} \sim$ 5 $\times$ 10$^{22}$ cm$^{-2}$.  The time-averaged mass transfer rate is related to the
time-averaged luminosity by $\dot{M}$ = $L$/$\eta c^{2}$, where $L$ is the time-averaged luminosity, $\eta$ is the accretion-luminosity efficiency (\citet{Shapiro1986} give $\eta$ = 0.1), and $c$ is the speed of light.  
This procedure yields $\dot{M} \sim$ 2 $\times$ 10$^{-10} \Msun/{\rm yr}$ for J1750, which, along with a quiescent luminosity of
 $(1.05 \pm 0.12)$ $\times$ 10$^{34}$ erg s$^{-1}$, falls outside of the range predicted by the cooling curves.

It is possible that the exceptionally high luminosity and NS surface temperature that we have observed for J1750 is due to the fact that the crust of the NS was not in thermal equilibrium with the core during our
observation, and that the crust has since cooled to a normal level.  This might occur if say, our observation took place too soon after an outburst.  The last known outburst to occur before 
our observation was in 2008 (see Figure 1) and persisted for $\sim$ 140 days \citep{Markwardt2008,Linares2008a}.  While monitoring J1750 as it decayed back down to quiescence with the {\em Swift} X-Ray Telescope (XRT), \citet{Linares2008b} detected 
a flare 67 days after the main burst event (This flare is denoted by the
triangle at MJD 54747 in the bottom panel of Figure 1).  Following this flare, {\em RXTE}/PCA continued to detect low-level activity until a data gap between MJD 54769$-$54870 prevented further coverage.  Once the monitoring was resumed 
on MJD 54870, {\em RXTE}/PCA continued to detect low-level activity for about 10 days, at which point the X-ray count rate had reached the baseline level.
Thus, the outburst may have extended out as far as MJD 54880.  {\em RXTE}/PCA did not detect J1750 between MJD 54880 and the time of the {\em XMM} observation (MJD 55293), and we estimate that the upper limit on the {\em RXTE}/PCA count rate is  
$<$ 5 counts s$^{-1}$ PCU$^{-1}$, which, using PIMMS with a $\Gamma$ = 2 power law, and with $N_{\rm H} =$ 6.0 $\times$ 10$^{22}$ cm$^{-2}$, gave a 2$-$10 keV luminosity upper limit of 2.7 $\times$ 10$^{35}$ erg s$^{-1}$.  This upper limit is low enough
that it is very likely that J1750 was in quiescence during this period of time.  We summarize by stating that J1750 spent at least 413 days in quiescence between
the 2008 outburst and our observation with {\em XMM}.  We note that two quasi-persistent sources, MXB 1659$-$29 and KS 1731$-$260,
were monitored repeatedly in X-rays after the cessation of outbursts lasting 2.5 yr and 12.5 yr, respectively \citep{Cackett2006}.  An exponential decay to a constant was fit to the 
cooling curves of these two sources, yielding  e-folding times of $325 \pm 101$ days for KS 1731$-$260 and $505 \pm 59$ days for MXB 1659$-$29 \citep{Cackett2006}. Therefore, given that J1750 is a transiently 
accreting system - i.e. the crust is likely not heated drastically out of equilibrium with the core during accretion episodes - and considering that the e-folding timescales for thermal relaxation 
in two quasi-persistent systems are comparable to the time spent in quiescence by J1750 before our observation, we suspect that the NS crust of J1750 was indeed close to its equilibrium 
temperature and luminosity during the observation.

Another possibility is that the quiescent luminosity of J1750 is variable and that we happened to observe it in a particularly luminous state.  This type of behavior has
been seen in two sources;  The quiescent luminosity of Cen X$-$4 was found to vary by a factor of 4.4
over a timescale of 7.5 years \citep{Cackett2010_x4}, which, as the authors show, requires variability in the thermal component.  \citet{Rutledge2002} performed multiple
observations of Aql X$-$1 in quiescence for five months following an outburst in 2000.  The luminosity was found to decrease by 50\% over the first 3 months, increase by 35\% over
the following month, and then remained constant over the final month.  While \citet{Rutledge2002} found that the variability could only be explained by a variable NS surface temperature, a
re-analysis by \citet{Campana2003} showed that the variability could be explained by correlated changes in the power law normalization and column density.  Furthermore, \citet{Cackett2011}
analyzed 10 more quiescent observations of Aql X$-$1 and could not conclude whether the power law component, the thermal component, or both components were responsible for the variability in
quiescent luminosity from epoch to epoch.

It is believed that the variability in quiescent luminosity is likely due to ongoing, low-level and/or episodic accretion during quiescence.  \citet{Fridriksson2011}
observed flaring behavior from XTE J1701$-$462 in quiescence during which the thermal flux, power law flux, and NS surface temperature were all found to increase significantly, resulting 
in a total luminosity of which $53 \pm 2$ ~\% was due to the power law component.  The authors attribute this to accretion episodes which overall are too faint to be detected by all-sky monitors.  
If a similar event had occured in J1750, one might expect to see a significant power law component in the spectrum.  But as reported in Section 3, we did not detect a power law component in the spectrum of J1750.  

The absence of a power law component in the spectrum of J1750 conflicts with the results of \citet{Jonker2004}, where an anticorrelation was found between the quiescent 0.5$-$10 keV luminosity $L_{\rm 0.5-10 keV}$ 
and the power law contribution to the luminosity.  The trend appears to reach a minimum at $\sim$ 1 - 2 $\times$ 10$^{33}$ erg s$^{-1}$.  At luminosities higher than this, the power law fraction appears to be correlated
with the luminosity as indicated by the multiple observations of Aql X$-$1 and XTE J1709$-$267 in quiescence.  The authors suggest that the power law component is due to residual accretion, an idea that is motivated by the
observation that the power law fraction of XTE J1709$-$267 was found to decrease as a function of time after the end of an outburst.  For J1750, we have $L_{\rm 0.5-10 keV}$ = $(8.8 \pm 1.2)$ $\times 10^{33}$ erg s$^{-1}$ and a maximum
power law fraction of 1.6\%, or 3.7\% if the black body fits are also considered.  This is in contrast to the correlation between power law fraction and luminosity, which would predict a power law fraction of 25$-$50\% for the observed
luminosity of J1750.  While J1750 does not appear to follow the trend of increasing power law fraction at higher quiescent luminosities, this area of parameter space
is rather sparse, and J1750 could be one of several unknown sources displaying such behavior.

\section{Conclusion}

Given the amount of time spent in quiescence before our observation, we conclude that J1750 was likely in an equilibrium state, with a NS surface temperature close to the core temperature.  The absence of a power
law component leads us to believe that the high luminosity was probably not due to an undetected accretion episode.  Follow-up observations of J1750 in quiescence could shed light on this conclusion, either by
detecting J1750 at a comparable luminosity and NS surface temperature and confirming it as an abnormal source, or by finding that it has cooled down to a relatively normal level consistent with NS cooling theory.

\acknowledgements

JAT and AB acknowledge partial support from NASA {\em XMM-Newton} Guest observer award number NNX09AP91G.  We thank Manuel Torres for useful discussions, and Craig Markwardt for making the PCA light curves easily accessible.  COH would like
to acknowledge NSERC funding.

\end{document}